\documentclass{article}
\usepackage{spconf,amsmath,graphicx}

\usepackage{graphicx}

\newcommand{\ie}{\textit{i}.\textit{e}.}

\usepackage{multirow,framed}
\usepackage{amsmath}
\usepackage{pifont}
\usepackage[normalem]{ulem}

\usepackage{amssymb}
\usepackage{latexsym}
\usepackage{url}
\usepackage{xcolor}
\usepackage{hyperref}
\usepackage{diagbox}
\usepackage{tabularx}
\usepackage{array}
\usepackage{ragged2e}
\usepackage{graphicx}
\usepackage{tikz}
\usepackage{bm}

\usepackage{enumitem}
\setlist{nosep, leftmargin=14pt}

\usepackage{mwe} 

\title{CP-Dilatation: A Copy-and-Paste Augmentation Method for Preserving the Boundary Context Information of Histopathology Images}
\name{Sungrae Hong$^{1}$ \space\space Sol Lee$^{1}$ \space\space Mun Yong Yi$^{1}$\sthanks{Corresponding Author}}
\address{$^{1}$Korea Advanced Institute of Science and Technology, Daejeon, South Korea}
\begin{document}
%
\maketitle
\begin{abstract}
Medical AI diagnosis including histopathology segmentation has derived benefits from the recent development of deep learning technology. However, deep learning itself requires a large amount of training data and the medical image segmentation masking, in particular, requires an extremely high cost due to the shortage of medical specialists. To mitigate this issue, we propose a new data augmentation method built upon the conventional Copy and Paste (CP) augmentation technique, called CP-Dilatation, and apply it to histopathology image segmentation. To the well-known traditional CP technique, the proposed method adds a dilation operation that can preserve the boundary context information of the malignancy, which is important in histopathological image diagnosis, as the boundary between the malignancy and its margin is mostly unclear and a significant context exists in the margin. 
In our experiments using histopathology benchmark datasets, the proposed method was found superior to the other state-of-the-art baselines chosen for comparison.
\end{abstract}

\section{Introduction}

With the rapid development of computing technology, deep neural networks (DNNs) are being applied to various practical tasks across industries. Notably, deep learning algorithms grounded in computer vision are actively applied to digital pathology~\cite{baig2020deep}. Histopathological image segmentation, in particular, is widely recognized as a valuable tool to aid pathologists in diagnosing conditions by evaluating the malignancy's spatial orientation and configuration~\cite{krithiga2021breast,zhao2022application}. 

\begin{figure}
 \centering\includegraphics[width=1.0\columnwidth]{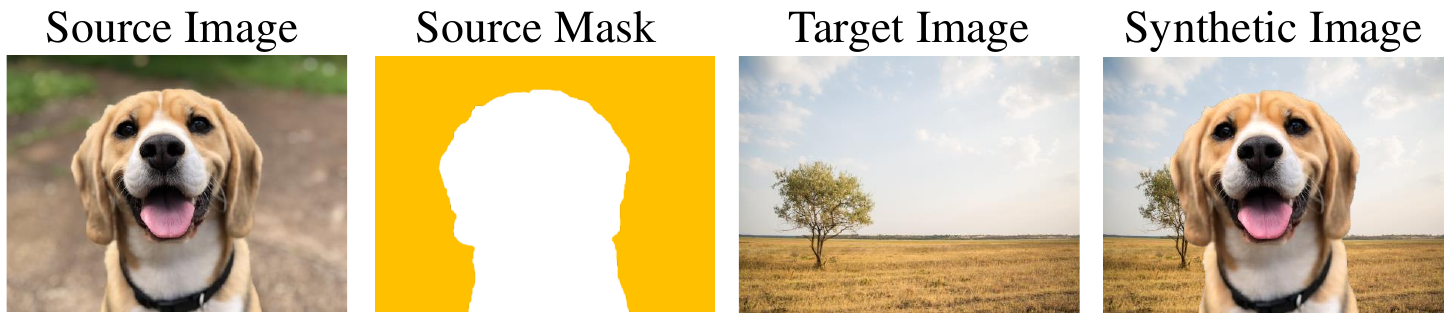}
 \caption[Constraints on segmentation image augmentation]%
{CP approach relies on the source mask associated with the source image, which indicates the precise location of the object to be copied. By leveraging this mask, we can isolate the object entirely from the source image and superimpose it seamlessly onto the target image.}
 \label{fig:1}
\end{figure}

While deep learning has been found useful in many applications, one serious limiting factor is its inherent demand of an extensive dataset to produce high quality representations~\cite{krizhevsky2017imagenet}. Building a huge dataset for deep learning takes a large amount of time and cost. From a task point of view, the labeling cost of segmentation requires time-consuming pixel-wise annotation. From a domain point of view, the cost of labeling medical datasets is very expensive as it often requires the assistance of medical doctors, and making things worse, it is constantly rising~\cite{kiryati2021dataset}, creating a serious conflict with the property of deep learning that consumes super-massive data.

Data augmentation serves as a solution to address the challenges posed by the high cost of data acquisition and the subsequent shortage of training data. This technique is well-known for enabling deep learning models to enhance their generalization performances by utilizing diverse, modified samples derived from existing data~\cite{fadaee2017data,fawzi2016adaptive}. In the segmentation tasks, the most commonly used data augmentation technique is Copy-and-Paste (CP), which involves copying an object from a source image and pasting it onto a target image to create a new image (as depicted in \autoref{fig:1}). Augmentation through CP has played a significant role in enhancing model performances and increasing the amounts of available segmentation data in recent research studies~\cite{fang2019instaboost,ghiasi2021simple}.

\begin{figure}
 \centering\includegraphics[width=1.0\columnwidth]{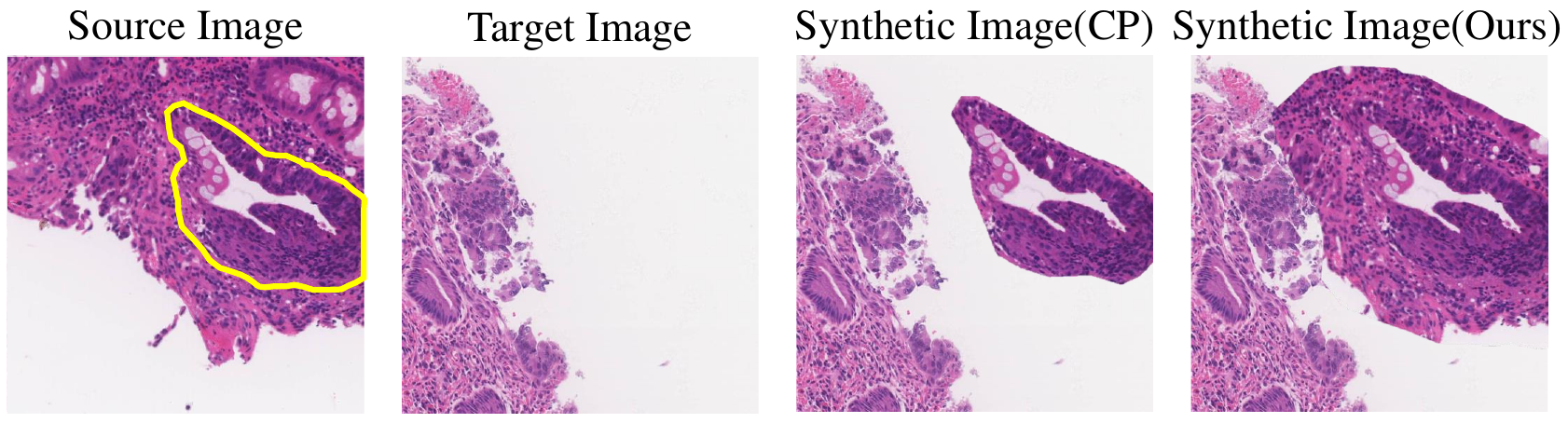}
 \caption[comparison between Naive CP and CP-Dilatation]%
 {The following examples illustrate the differences between naive CP and our {CP-Dilatation}. Naive CP copies the malignancy from the source image and pastes it onto the target image, thus limiting the utilization of information beyond the annotated region. Conversely, {CP-Dilatation} copies the context outside of the malignancy as indicated by the source mask and pastes it onto the target image. As a result, the image augmented with {CP-Dilatation} captures both the malignancy and its boundary information.}
 \label{fig:2}
\end{figure}

However, prior CP research has not adequately addressed the issue of augmenting histopathology images that feature complex contexts surrounding malignancy~\cite{sun2019comparative}. It is important to recognize that the boundary context of a malignancy is just as significant as the malignancy itself~\cite{fleming2012colorectal}, as the expression of the malignancy is influenced by the surrounding area~\cite{decostanzo1997necrosis,okuyama2003budding}. Properly examining the periphery can, therefore, aid in the diagnosis.


To overcome the limitations of prior research, we present a new augmentation technique, {CP-Dilatation}, that allows for the contextual information surrounding malignancies in histopathology segmentation images to be captured, while also increasing the size of available data. By applying a dilatation operation to the mask used to extract malignancies from the source image, in this study we demonstrate that we can properly retrieve the surrounding contextual information, as illustrated in \autoref{fig:2}. While dilatation is a common method used for reducing image noises, we have found that it can also be utilized to take advantage of the contextual details surrounding malignancies. To the best of our knowledge, this is the first CP methodology study solely devoted to histopathological images. We evaluated the performance of our methodology using the DigestPath2019~\cite{da2022digestpath} and Warwick-QU~\cite{sirinukunwattana2017gland} Whole Slide Image (WSI) benchmark datasets, and our experiments indicate that our proposed approach leads to performance improvements across various metrics. We summarize the contributions of our study as follows.

\begin{enumerate}
\item {We propose {CP-Dilatation} that is tailored to the specific characteristics of histopathological images. As far as we know, this is the first CP methodology designed for histopathological images and their malignancies.}


\item {The performance of our proposed method and a comparative methodology was evaluated objectively using publicly available datasets. Our methodology demonstrated superior performances compared to the comparative methodology.}

\end{enumerate}
\section{Method}


\begin{multline}\label{eq:1}
X^{aug}_{src}, X^{aug}_{tar}, M^{aug}_{src}, M^{aug}_{tar}\\
=\mathrm{Aug}_{spatial}(\mathrm{Aug}_{optical}(X_{src},X_{tar}),M_{src},M_{tar})
\end{multline}

\begin{subequations}\label{eq:2}
\renewcommand{\theequation}{\theparentequation.\arabic{equation}}
\begin{align}
&M_{src_{D}}=M^{aug}_{src}\otimes{K}\label{eq:2-1}\\
&O_{src} = X^{aug}_{src}\odot{M_{src_{D}}}\label{eq:2-2}
\end{align}
\end{subequations}

\begin{subequations}
\renewcommand{\theequation}{\theparentequation.\arabic{equation}}
\begin{align}
\begin{split}
&{X_{new}}=\left\{{(1-M_{src_{D}}})\odot{X^{aug}_{tar}}\right\}\oplus\left\{\mathrm{B}(M_{src_{D}})\odot{O_{src}} \right\}\label{eq:3-1}\\
\end{split}\\
\begin{split}
&M_{new}=\min(\max(M^{aug}_{tar}\ominus{M_{src_{D}}},0)\oplus{M^{aug}_{src}},1)\label{eq:3-2}
\end{split}
\end{align}
\end{subequations}


\begin{figure}
    \centering\includegraphics[width=1.0\columnwidth]{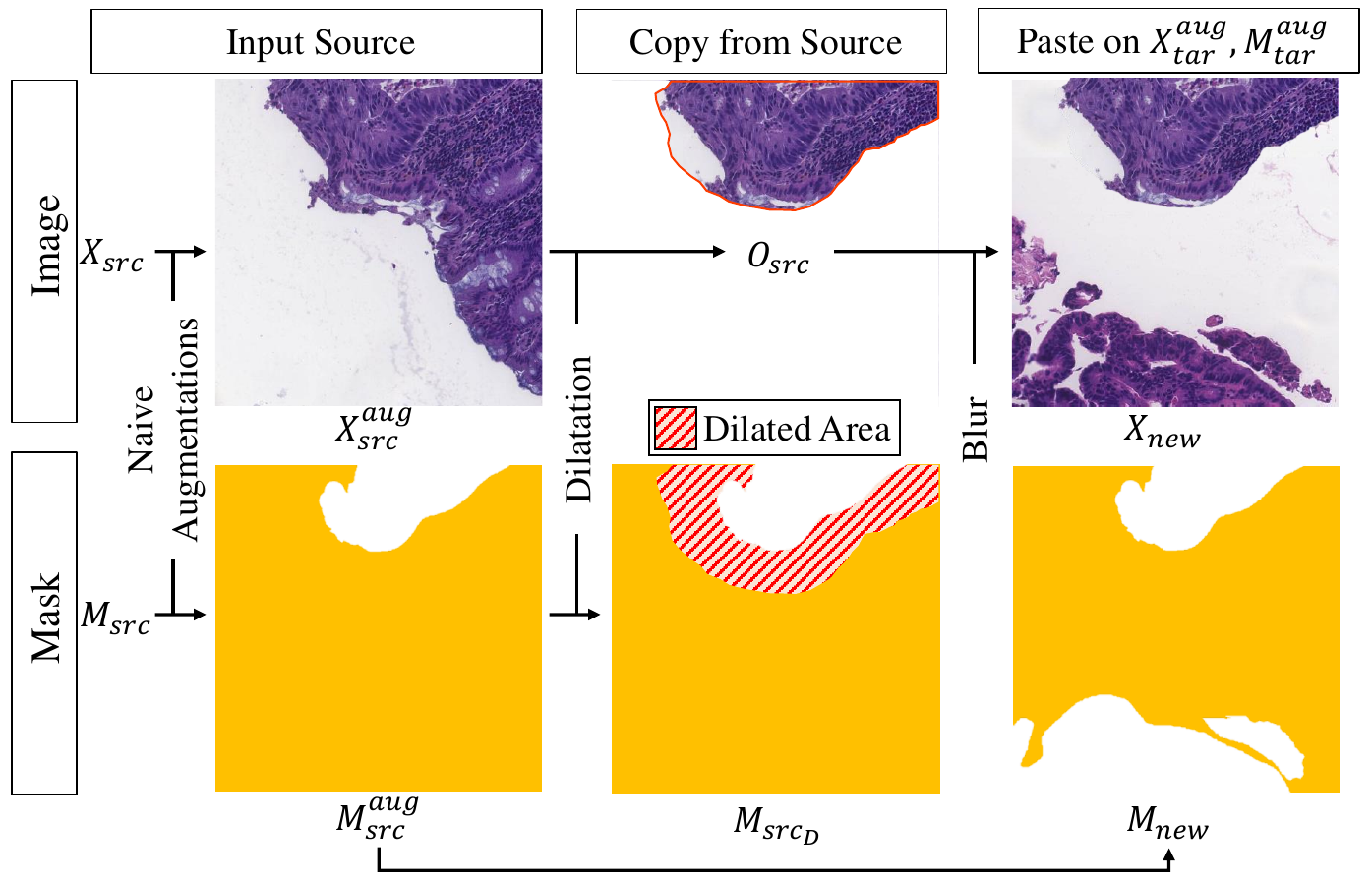}
    \caption[]%
{{CP-Dilatation} pipeline. The inputs are ($X_{src}$, $X_{tar}$) and ($M_{src}$, $M_{tar}$), which are the images and masks of the source and target, respectively. The dilatation operation, which makes $M^{aug}_{src}$ more plump, produces $M_{src_{D}}$. The red dashed line in figure $M_{src_{D}}$ shows this. As a result, $M_{src_{D}}$ makes it possible to copy not only the malignancy from $X^{aug}_{src}$ but also the context around it.}
    \label{fig:whole}
\end{figure}

{CP-Dilatation} consists largely of three steps. First, conventional augmentations change the spatial and optical information of the source and target. Second, the Copy Step makes a chubby mask through a dilatation operation and the source image's object containing malignancy and its surrounding context. For the last, Paste Step outputs the synthesized results $X_{new}, M_{new}$. Note that $\odot, \oplus, \ominus$, and $\otimes$ represent element-wise multiplication, element-wise sum, element-wise subtraction, and convolution operations, respectively.

\subsection{Naive augmentations}\label{augmentations}
Naive augmentations increase data diversity and model robustness~\cite{fawzi2016adaptive}. we categorized naive augmentations into two distinct categories: spatial and optical. We applied \textit{VerticalFlip, HorizontalFlip}, \textit{Random90-DegreeRotation}, and \textit{RandomResizePad}(-0.9,0.5) as spatial augmentations. We used \textit{RandomContrast} and \textit{RandomBrightness} as optical augmentations. We applied each naive augmentation to the sample with the same probability $p_{aug}$ to the source and target. \autoref{eq:1} express this process.

\subsection{Copy Step}\label{copy}

The purpose of the Copy Step is to extract malignancy and the meaningful context surrounding it together from a given source image $X^{aug}_{src}$ and mask $M^{aug}_{src}$.

\begin{figure}
 \centering\includegraphics[width=1.0\columnwidth]{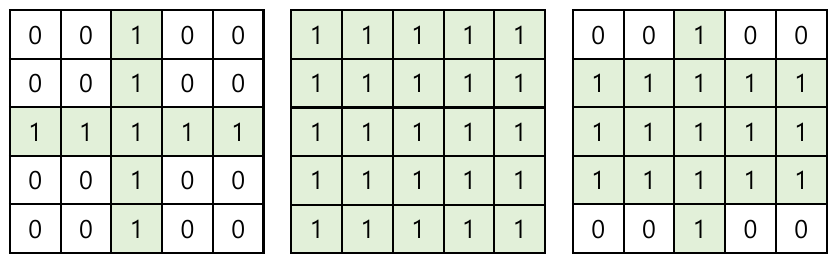}
 \caption[Various kernels]%
 {We use various kinds of kernels for dilatation operations. From the left are $5\times{5}$ \textit{Cross(Dilatation), Rectangular}, and \textit{Open} kernel respectively. The kernels are filled with values indicating how much to refer to the outside pixel based on the center. The more $1$ in the kernel, the more surrounding context will be used, and the more chubby the mask will be.}
 \label{fig:99}
\end{figure}

Dilatation or dilation is a basic convolution operation for removing noise from an image. This operation spreads values around a boundary pixel to neighborhood pixels, so the result is more chubby than the original. We note that if this operation is applied to the binary mask, the binary value can be extended beyond the annotation. This quick and simple convolution operation can create a mask that does not discard context near the malignancy in the histopathological images.
We used various kernels presented in \autoref{fig:99}. We obtained dilated source mask $M_{src_{D}}$ by convolution operation of this kernel $K$ to $M^{aug}_{src}$(\autoref{eq:2-1}).

$M_{src_{D}}$ can be considered an annotated mask of malignancy and its surroundings. Thus, we obtain the copied object $O_{src}$ containing malignancy of the source image and their surroundings by \autoref{eq:2-2}.

\subsection{Paste Step}\label{paste}
Mixing two or more samples requires jointing heterogeneous information. In particular, the heterogeneity of this information stands out at the boundary where the source image and the target image meet. Therefore, We used a Gaussian blur, which mitigates heterogeneous information, to $M_{src_{D}}$. This corresponds to $\mathrm{B}(\cdot)$. We computed this with $O_{src}$ using $\odot$ operation to obtain the blurred $O_{src}$. The $(1-M_{src_{D}})\odot{X^{aug}_{tar}}$ operation in \autoref{eq:3-1} wipes off only the source image portion to be synthesized from the target image. We obtained a new synthesized image $X_{new}$ by $\oplus$ operation of all of the above results.

$M_{src_{D}}$ is a mask or alpha that extends the actual malignancy and its surroundings. Therefore, since $O_{src}$ is made from $M_{src_{D}}$, it partially obscures the context of target image $X^{aug}_{tar}$. So we use a different mask synthesis method from the conventional CP methodologies. In image synthesis, the overlapping portion of $O_{src}$ and $X^{aug}_{tar}$ is deleted from $X^{aug}_{tar}$. So we subtracted $M_{src_{D}}$ from $M^{aug}_{tar}$. Here, to prevent this output from becoming $-1$({\ie} When the dilated area erases the background of the target.), we perform $\max(\cdot)$ operation with $0$. We add the actual malignancy annotation in $O_{src}$ using $\oplus$ operation of $M^{aug}_{tar}$ with the previous output. At this time, to prevent this being $2$({\ie} The malignancy in the source and target is situated at overlapping positions.), we perform $\min(\cdot)$ operation with $1$. The result of these operations is a synthesized mask $M_{new}$(\autoref{eq:3-2}).
\section{Experiment}

\begin{table*}
\centering
\caption{Comparison results with proposed methodology}.
\label{tab:comparison}
\resizebox{0.9\textwidth}{!}{%
\begin{tabular}{l|ccc|ccc|ccc|ccc} 
\hline
\multicolumn{1}{c|}{\multirow{3}{*}{-}} & \multicolumn{6}{c|}{DigestPath2019\cite{da2022digestpath}}                                         & \multicolumn{6}{c}{Warwick-QU\cite{sirinukunwattana2017gland}}\\ 
\cline{2-13}
\multicolumn{1}{c|}{}                   & \multicolumn{3}{c|}{Unet\cite{ronneberger2015u}} & \multicolumn{3}{c|}{DeepLab v3+\cite{chen2018encoder}} & \multicolumn{3}{c|}{Unet} & \multicolumn{3}{c}{DeepLab v3+}\\ 
\cline{2-13}
\multicolumn{1}{c|}{}                   & Dice           & IoU            & PA             & Dice           & IoU            & PA             & Dice          & IoU              & PA       & Dice  & IoU   & PA \\ 
\hline
{None}                           & 0.621          & 0.490          & 0.916          & 0.631          & 0.501          & 0.915          & 0.871         & 0.773            & 0.934     & 0.866 & 0.772 & 0.932         \\
{Naive Augmentations}            & 0.644          & 0.517          & 0.920          & 0.633          & 0.509          & 0.918          & 0.876         & 0.782            & 0.938     & 0.895 & 0.812 & 0.942            \\
{CP-Naive}                       & 0.672          & 0.547          & 0.926          & 0.658          & 0.535          & 0.926          & 0.845         & 0.735            & 0.925     & 0.852 & 0.745 & 0.927            \\
{CP-Naive w. Gaussian}           & 0.670          & 0.544          & 0.926          & 0.664          & 0.534          & 0.925          & 0.847         & 0.741            & 0.927     & 0.852 & 0.748 & 0.929            \\
{CP-Simple}\cite{ghiasi2021simple}& 0.667         & 0.544          & 0.925          & 0.621          & 0.497          & 0.914          & 0.889         & 0.801            & 0.941     & 0.890 & 0.804 & 0.943            \\
{TumorCP}\cite{yang2021tumorcp}  & 0.656          & 0.529          & 0.922          & 0.645          & 0.518          & 0.920          & 0.869         & 0.770            & 0.930     & 0.863 & 0.762 & 0.929            \\ 
\hline\hline
{CP-Dilatation(DILATE-10-0.4)}            & \textbf{0.687} & \textbf{0.564} & \textbf{0.928} & 0.666          & 0.543          & \textbf{0.927} & 0.897         & 0.815            & 0.943     & 0.904 & 0.826 & \textbf{0.947} \\
{CP-Dilatation(RECT-30-0.7)}              & 0.676          & 0.552          & 0.927          & \textbf{0.671} & \textbf{0.548} & \textbf{0.927} & 0.897         & 0.814            & 0.944     & \textbf{0.905} & \textbf{0.827} & \textbf{0.947} \\
{CP-Dilatation(OPEN-40-0.0)}              & 0.675          & 0.550          & 0.924          & 0.667          & 0.543          & 0.926          & \textbf{0.901}& \textbf{0.821}   & \textbf{0.945} & 0.903  & 0.825 & \textbf{0.947}  \\
\hline
\end{tabular}
}
\end{table*}

\subsection{Experimental settings}

\noindent\textbf{Implementation details}
We leveraged the Resnet34~\cite{he2016deep} as the backbone network. The probability assignments for naive augmentation $p_{aug}$ and {CP-Dilatation} were set at $0.33$ and $0.5$, respectively. However, a $p_{aug}$ value of 1 was exclusively applied to \textit{RandomResizePad}. Our optimization strategy involved using the Adam optimizer~\cite{kingma2014adam} with $\bm{\beta}=[0.9,0.999]$, employing binary cross-entropy alongside the dice loss as the objective functions. We evaluated our model's malignancy segmentation performance in pixel units, based on Dice Coefficient(Dice), Intersection of Union(IoU), and Pixel Accuracy(PA). All hyperparameters of {CP-Dilation} were chosen based on the validation set. We report the average performance by conducting multiple repetitions of experiments.

\par\smallskip
\noindent\textbf{Dataset}
We used DigestPath2019~\cite{da2022digestpath} by cutting them into patch unit images. The DigestPath2019 is high-capacity data in gigapixel units. Therefore, We truncated all images to size 1024 by 1024 and deleted several images if more than 75\% of pixels are in the range HSV space criteria (0, 0, 255×0.9),(360, 255×0.12, 255). We established the train/validation/test sets at proportions of 0.7, 0.15, and 0.15, respectively. The splitting was performed at the patient level to avoid cheating. We also used Warwick-QU~\cite{sirinukunwattana2017gland} for experiments. Both datasets were resized to 512 pixels.

\par\smallskip
\noindent\textbf{Baseline}
To establish an objective measure of performance, we have pitted {CP-Dilatation} against various baseline methodologies. {CP-naive} is a simplistic approach that entails copying and pasting the source image's object onto the target. Alternatively, we have also experimented with {CP-naive} with Gaussian blur, a commonly used technique to enhance its performance. {CP-simple}~\cite{ghiasi2021simple}, a CP-based methodology that has demonstrated state-of-the-art performance on the COCO dataset~\cite{lin2014microsoft} for natural images, is also employed as a baseline.
{TumorCP}~\cite{yang2021tumorcp}, a CP-based methodology for kidney tumors, stands out as the first of its kind proposed for medical domains, as per their assertion.

\subsection{Comparison results}
\autoref{tab:comparison} shows the performance of the baseline and our proposed methodology. {None} is the result of training the model without any augmentations to the data. {Naive Augmentation} consists of elements proposed in \ref{augmentations}. We suggest parameters of {CP-Dilatation} as (kernel type-kernel size-$\sigma$). In general, {Naive Augmentation} improved the performance of {None}. In the case of {CP-Naive} and {tumor-CP}, it contributed to the performance enhancement in DigestPath2019, but not in Warwick. This is attributed to the nature of the more zoomed-in Warwick, where simple copy-paste obscured crucial information for learning. As evidence, {CP-Simple} and {CP-Dilatation}, which included object resizing, consistently demonstrated high performance on both datasets. Especially, our methodology outperformed the baseline on two datasets and models.

\begin{figure}
 \centering\includegraphics[width=0.9\columnwidth]{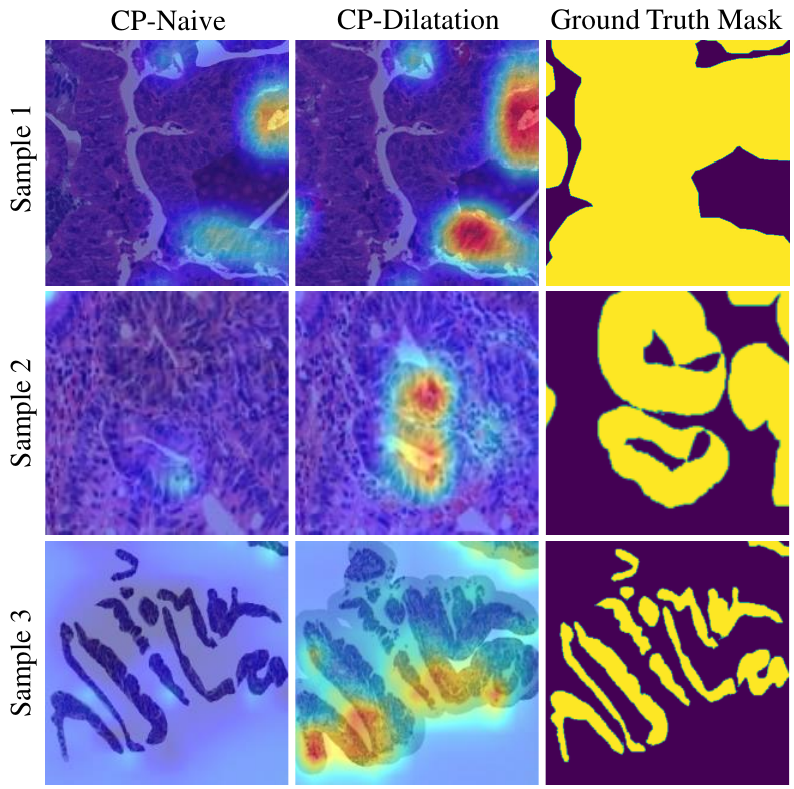}
 \caption[comparison between Naive CP and CP-Dilatation]%
 {GradCAM of the synthesized samples with {CP-Naive} and {CP-Dilatation}.}
 \label{fig:5}
\end{figure}

\subsection{Post-hoc analysis}
\noindent\textbf{Experiment design}
\,The purpose of the experiment is to qualitatively verify whether providing the dilated area to the deep neural network aids in concentrating better on the malignancy. We trained ResNet18~\cite{he2016deep} classifier on DigestPath2019. Subsequently, we inferred this model using two synthesized DigestPath2019 datasets created through CP-Naive and CP-Dilatation respectively. Both methods homogenized all conditions except for the presence of the dilatation operation. GradCAM~\cite{selvaraju2017grad} visualizes where in the input image the network relied on to make predictions.

\par\smallskip
\noindent\textbf{Result}
\:\autoref{fig:5} shows the output of Gradient CAM. The neural network showed a closer concentration of the ground truth malignancies in the image generated by {CP-Dilatation} compared to those generated by {CP-Naive}. It indicates that providing slightly more information around the malignancies can help the neural network make a more definitive diagnosis.
\section{Conclusion}
We present an novel augmentation method that tackles the pervasive problems in the histopathology image segmentation tasks: high labeling cost and the scarcity of training data. In this study, we have conceived that valuable contexts can exist beyond malignancy, and thus devised a tailored dilatation operation to extract them. Our approach is a pioneering effort in extracting helpful contextual information with mask dilatation and represents the first attempt to augment histopathology segmentation image data. Through experiments, we have demonstrated the effectiveness of our methodology in accurately segmenting malignancy. Our findings contribute to solving the challenge of data scarcity in histopathology segmentation and opening up new possibilities for coping with labeling cost issues in digital pathology.

\par\smallskip
\noindent\textbf{Compliance with Ethical Standards.}
This research study was conducted retrospectively using human subject data made available in open access. Ethical approval was not required as confirmed by the license attached with the open access data.

\end{document}